\documentclass[twocolumn,a4paper,floatfix]{revtex4} 
\usepackage{graphicx,amssymb,amsfonts,amsmath,float}

\newcommand{\beq}{\begin{equation}}
\newcommand{\eeq}{\end{equation}}
\newcommand{\bea}{\begin{eqnarray}}
\newcommand{\eea}{\end{eqnarray}}
\newcommand{\no}{\nonumber}
\newcommand{\gt}{\tilde{g}}
\newcommand{\scale}{
\begin{figure}[ht]
\begin{center}
\includegraphics[width=80mm, clip]{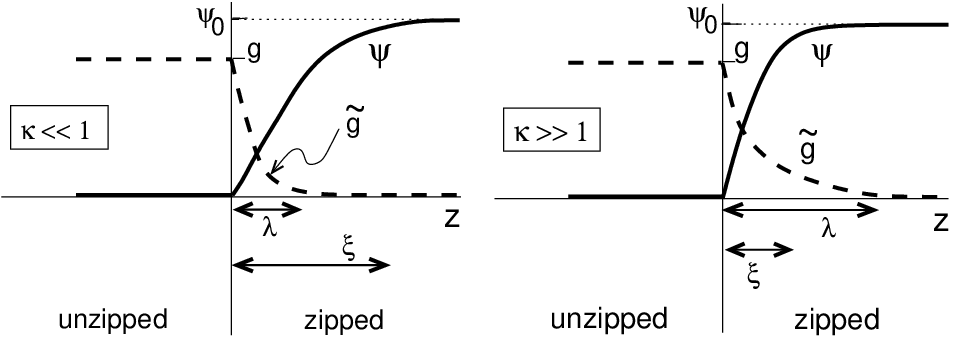}
\caption{
Schematic diagram of variation of zipping-unzipping order
  parameter $\psi$ (continuous line) and applied force $g$ (dashed
  line) inside unzipped and zipped phases. $\xi$ is the length scale of
  variation of $\psi$ inside the zipped phase and $\lambda$ is the scale
  for $g$. For Type I (left figure), $\kappa=\lambda/\xi \ll 1$ and for
  Type II (right figure), $\kappa \gg 1$.  }
\label{fig.1}
\end{center}
\end{figure}
}
{
\newcommand{\defect}{
\begin{figure}[ht]
\begin{center}
\includegraphics[width=80mm,clip]{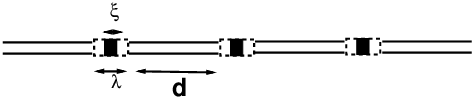}           
\caption{Schematic diagram of a periodic array of defect blobs.
The array
  has a periodicity $d$ which controls the density of blobs.  Each distorted
  region is of length $\sim \lambda$ with an unzipped core of size $\sim
  \xi$.
  }
\label{fig.2}
\end{center}
\end{figure}
}

\begin{document}
\title{TYPE II DNA: when the interfacial energy becomes negative}
\author{Poulomi Sadhukhan}
\email{poulomi@iopb.res.in}
\author{Jaya Maji}
\email{jayamaji@iopb.res.in}
\author{Somendra M. Bhattacharjee}
\email{somen@iopb.res.in}
\affiliation{Institute of Physics, Bhubaneswar 751 005, India}

\begin{abstract} 
  An important step in transcription of a DNA base sequence to a
  protein is the initiation from the exact starting point, called
  promoter region. We propose a physical mechanism for identification
  of the promoter region, which relies on a new classification of DNAs
  into two types, Type-I and Type-II, like superconductors, depending
  on the sign of the energy of the interface separating the zipped and
  the unzipped phases. This is determined by the energies of helical
  ordering and stretching over two independent length scales. The
  negative interfacial energy in Type II DNA leads to domains of
  helically ordered state separated by defect regions, or blobs,
  enclosed by the interfaces. The defect blobs, pinned by non-coding
  promoter regions, would be physically distinct from all other types
  of bubbles. We also show that the order of the melting transition
  under a force is different for Type I and Type II.
\end{abstract}

\maketitle

DNA in its double helical form shows a resilience against an external
pulling force.  The bound state does not allow a force $g$ applied at
an end to penetrate up to a critical force $g=g_c$, above which the
DNA gets unzipped\cite{smb99,unzip01,kumar}.  The transition is first
order for temperatures $T<T_c$ where $T_c$ is the
denaturation(melting) temperature in the absence of any
force\cite{gotoh}.  The force-induced unzipping transition of DNA is
due to a competition between the bond orientation by force and
ordering by base pairing. The formation of a helically ordered dsDNA
from denatured strands is a symmetry breaking transition. At a
coarse-grained level, the ordered state can be described by an order
parameter $\psi$, with $\psi=0$ for the denatured state. The external
force does not couple directly to this order parameter. Consequently,
at a junction of a bound and an unzipped DNA, there is a need to
define two length scales: one scale $\xi$ that gives the length over
which the DNA ordering is damaged on the bound side of the interface,
while the other scale $\lambda$ gives the distance over which the
force penetrates the bound state.  The existence of the second scale
$\lambda$ was pointed out by de Gennes in a model involving stretching
of the backbone and the hydrogen bonds\cite{degen}. Generally one
expects interfaces separating phases to be energetically costly (e.g.
surface tension), but here we show that if $\lambda\gg\xi$, then the
interfacial energy, or surface energy, between bound and unzipped DNA
can become negative.  There can then be a penetration of force in the
form of distorted regions or ``defect blobs'' of length $\lambda$
enclosing a denatured bubble of size $\xi$.  In analogy to
superconductors, when the interfacial energy becomes negative, one
gets a mixed phase of DNA and the zipped-mixed phase transition
becomes continuous. Based on the sign of the zipped-unzipped
interfacial energy we classify DNA into two types: Type II has
negative interfacial energy whereas Type I is the conventional case
with positive interfacial energy. This classification is not related
to the existing classification based on DNA conformation.

A Type II DNA has novel features which are of considerable biological
and physical implications. To be noted that the defect blobs are
different from thermally created bubbles.  This is because the bubbles
of the latter type would consist of random configurations of denatured
strands generated by thermal fluctuations and may have positive
interfacial energy.  The distinctness of the defect blobs can be a
signature for their identification in biological processes.  Let us
consider the transcription process where the genetic code, determined
by the base sequence, is transferred to the amino acid sequence of a
protein. For correct transcription, the sequence must be read from the
correct starting point on DNA. These starting non-coding regions are
called promoter regions and their identification is the first and
vital step in transcription\cite{watson}.  A pulling force or a forced
separation in a homogeneous Type II DNA produces a finite density of
the defect blobs\cite{goodman} (discussed later).  The noncoding
sequences or the promoter regions may act as inhomogeneities on a DNA
and could play the role of pinning centers for the defect blobs.  The
advantage of physical identification of pinned defect blobs could
facilitate recognition of the promoter regions for gene expression
(e.g. see~\cite{cao,choi}). So far as physical properties are
concerned, Type I and Type II DNA will have different phase diagram
and phase transition as discussed later.

Recently, both in experiment\cite{mameren} and
simulation\cite{maren10}, a continuous transition has been observed if
the topology is preserved in a stretching experiment by pulling both
the strands either at both ends or at one end of an anchored DNA.  We
also note that a detailed molecular dynamics study\cite{randall} of
under- or over-wound DNA without writhe, observed the formation of
localized sequence-dependent defects which allow the rest of the dsDNA
to be in the relaxed normal state.  It is known that topoisomerase II
may bind anywhere on the DNA but its topology changing activity is
restricted to specific sequences (cleavage sites) indicating that
geometric distortions get localized around certain
sequences\cite{planitz}.  These are consistent with our general
predictions, though we like to add that interfacial information in any
of these cases are not available.

The thermodynamic description of unzipping of DNA requires three
variables, $\psi$ describing the helical ordering (i.e., broken
symmetry) and a force-displacement $(g,x)$ conjugate pair, where $x$
is the scaled separation between the two strands at the point of
application of force $g$.  On the bound side $x$ can be taken as the
response to an internal induced force $\tilde{g}$, so that,
\beq
x(\tilde{g})=\chi \tilde{g},
\eeq
where $\chi$ the stretchability or the
response function, is independent of $g$ in the linear response regime.
Though we restrict to linear response regime here, the final results
can be reproduced for a general force-dependent $\chi$. The variables
are chosen such that $\psi=0$ for the unzipped state, and $\psi\ne 0$
for the ordered state, while $\tilde{g}=0$ in the bulk of
the ordered state. At this point it is to be noted that the order
parameter $\psi$ represents helical ordering which is not directly coupled to
the external pulling force. As a result we get two independent length
scales in the problem. This makes the present treatment different from
other existing models.
 
For a homogeneous state, the Gibbs free energy $G(T,g)$ per unit
length at temperature $T$ and a pulling force $g$ is given by
\begin{equation}
\label{eq:3}
G(T,g) = G(T,0)- W(g), 
\end{equation}
where $W(g)=\int_0^g x(g') \; dg'$
is the work for stretching.  The conditions of
phase coexistence at $g=g_c$, {\it i.e.}
\begin{equation}
  \label{eq:5}
 G_{\sf z}(T,g_c) = G_{\sf u}(T,g_c), 
\end{equation}
and non-penetration of force in the bound state for $g\le g_c$, {\it
  i.e.},
\begin{equation}
  \label{eq:6}
  G_{\sf z}(T,g) = G_{\sf z}(T,0),
\end{equation}
(subscripts z and u representing the zipped and the unzipped phases),
when combined with Eq.~\ref{eq:3}, give
\begin{equation}
  \label{eq:4}
 G_{\sf z}(T,g) = G_{\sf u}(T,g)+W(g) - W(g_c).
\end{equation}
Eq.~\ref{eq:4} agrees with the known exact results of Ref.
\cite{unzip01} when appropriate $x(g)$ from the exact solution is
used. In particular one verifies that $G_{\sf z} - G_{\sf u} =
\frac{1}{2} \chi (g^2-g_c^2),$ in the linear response regime (near
melting).

Compared to the stretched unzipped state, the zipped phase has to pay
a cost $W(g)$ for force expulsion for not following the force-diktat,
but gains energy $W(g_c)$ due to binding or ordering.  The phase
coexistence requires a perfect compensation of one by the other.  This
compensation may be used to obtain the binding energy of the zipped
phase as
\begin{equation}
  \label{eq:7}
E_{\sf  z}(T)=W(g_c).  
\end{equation}
This equation may also be used to define $g_c$ from the binding energy.

Let us now consider an inhomogeneous situation of a dsDNA at $T<T_c$
by pulling at one end by a force $g=g_c(T)$ so that there is an
interface separating the coexisting zipped and unzipped phases.  The
interfacial energy is obtained by comparing this mixed state free
energy with that of a fully unzipped homogeneous state at $g=g_c$.
Needless to say that an interface can be created spontaneously if
there is a gain in energy in doing so.

Since far from the interface, the Gibbs free energy density is the
same in the two phases, the total free energy ${\cal G}$ can be
written as
\begin{equation}
\label{eq:11}
{\cal G}=\int_{-\infty}^{\infty} G_{\sf u}(T,g_c) dz +\sigma, 
\end{equation}
where $\sigma$ is the ``surface tension'', and $z$ is a contour length
measured along the DNA or the strands, the $z=0$ point being chosen at
the point of interface with $z<0$ as the unzipped phase.

We start with the free energy  functional
\begin{equation}
  \label{eq:8}
{\sf  F}_{\rm{tot}}=\int_{-\infty}^{\infty} dz\; {\cal F}\{\psi,x\},  
\end{equation}
whose minimum gives the equilibrium free energy in a fixed distance
ensemble.  The functional ${\cal F}\{\psi,x\}$ can be taken as
\begin{eqnarray}
{\cal F}\{\psi, x\} &=&F_{\sf u}+F\{\psi\}+\frac{K_\psi}{2}\left( \frac{\partial \psi}{\partial z}\right)^2 + \no\\
&~&\frac{K_x}{2}\left( \frac{\partial x}{\partial z}\right)
^2 + \int_0^{x}  g(\tilde{x})~ d\tilde{x}.\label{eq:1} 
\end{eqnarray} 
where $F\{\psi\}$ is the free energy of the homogeneous bulk zipped
phase with reference to the unzipped state free energy $F_{\sf u}$.
In the unzipped state $F\{\psi\}=0$.  $K_{\psi}$ and $K_x$ are
additional ``elastic'' constants for distortions in $\psi$ and $x$.
The elastic part of the free energy can be extended to torques.  The
order parameter $\psi$ and force $\gt$ are not coupled in the free
energy in the form taken in Eq.~\ref{eq:1} and consequently, this form
is valid only in extreme limits. Further generalizations are not
needed for this paper.  The Gibbs free energy is obtained from
Eq.~\ref{eq:1} by using the equilibrium values of $\psi$ and $x$,
followed by a Legendre transformation from $x$ to $g$.

\scale

The equilibrium conditions, obtained by minimizing ${\sf
  F}_{\rm{tot}}$, are \bea \frac{\delta F}{\delta \psi}-
K_{\psi}\frac{\partial ^2\psi}{\partial
  z^2}&=&0, \label{eq:gl1}\\
-K_x\frac{\partial ^2 x}{\partial
  z^2}+\frac{x}{\chi}&=&0,\label{eq:gl2} \eea with the condition that
\begin{equation}
  \label{eq:9}
  \psi=0,\ \ \ \ x=x_c=\chi g_c\ \ \ \ {\rm at}\ \ \ \ z=0,
\end{equation}
and
\begin{equation}
  \label{eq:10}
\psi=\psi_0,\ \ \ \ x=0\ \ \ \ {\rm at} \ \ \ \ z\to\infty,  
\end{equation}
$\psi_0$ being the solution of $$\frac{\delta F}{\delta \psi}=0$$ to
maximize the interfacial energy.  The length scales $\xi$ and
$\lambda$, giving how fast $\psi$ or $\gt$ grow or decay inside the
zipped phase (see fig.~\ref{fig.1}), come from Eqs. \ref{eq:gl1} and
\ref{eq:gl2}, as
\begin{equation}
  \label{eq:2}
  \xi^{-2}=\frac{1}{K_{\psi}} \left.\left( \frac{1}{\psi}\;\frac{\partial
      F}{\partial \psi}\right)\right|_{\psi\to 0},\ {\rm and\ }
\lambda^{2} = K_{x}\chi.
\end{equation}
The equation for $\lambda$ reduces to the form derived by de
Gennes~\cite{degen} if the elastic constants of his model are used for
$K_x$ and $\chi$.  The dimensionless ratio $\kappa=\lambda/\xi$ is
expected to be different for different sequences of DNA.

For $\kappa\ll 1$, the external force penetrates only a short distance
$\lambda$ into the zipped region. In contrast the order parameter
rises to its asymptotic value $\psi_0$ in a much larger length $\xi$.
One has to pay the energy cost for the damage in ordering over a
length scale $\xi$, and therefore, 
\beq 
\sigma \sim E_{\sf z} \xi \sim\frac{1}{2}\chi g_c^2\xi 
\eeq 
which is positive. This is the conventional scenario of force
expulsion of various models on the zipping-unzipping phase transition
and this scenario gives the well-known behavior of the unzipping
transition.

When $\kappa\gg 1$, the force penetrates a greater distance $\lambda$
into the sample, so that there is an obvious gain in the stretching
energy (i.e. reduction of the ``positive energy'' for force expulsion)
over the interval of penetration, over and above the gain by ordering.
With $x$ from Eq.~\ref{eq:gl2}, Eq.~(\ref{eq:11}) gives \beq \sigma
=-\frac{\chi g_c ^2}{2}\lambda, \eeq which is negative.  Hence, it is
possible to lower the free energy of the DNA by creating the
interface. The value of $\kappa$ for transition from Type I to Type II
depends on the form of $\chi$ which, in turn, depends on the DNA
sequence and the secondary structure.  It is therefore primarily the
sequence but also the secondary structure that determine whether a DNA
would behave like Type I or II.

\defect

If we now consider the bulk zipped state with a force $g$, then force
penetration may be possible in the form of many isolated distorted
regions or blobs.  For $\lambda \gg \xi$, with the unzipped core of
size $\xi$ costing an energy $E_{\sf z}\xi$, and the $x$ part of the
free energy ${\cal F}\{ \psi,x\}$ in Eq.~\ref{eq:1}, one finds for a
homogeneous chain that a periodic structure of the
blobs\cite{goodman}, as in fig.~\ref{fig.2}, is possible
energetically, if $g>g_c/\sqrt{\kappa}$.  The initial penetration of
force is at $g_{c1}=g_c/\sqrt{\kappa}$ with periodicity $d\to\infty$.
The unzipping transition therefore becomes continuous in contrast to
the first order nature for Type I.

The negative interfacial energy is found in Type II
superconductor\cite{abrikosov} too. Our formulation is similar to that
of Type II superconductivity in a one-dimensional geometry. As there
is indeed a phase transition in DNA, the Landau theory is justified
here. It suffices for a one dimensional case to consider a scalar
order parameter.

We may point out a few additional implications of a negative
interfacial energy. The penetration of the force is not possible in
the conventional polymer models.  For any helical or twisted pairs of
strings, a pulling force produces over-winding.  We expect this
over-winding in DNA to be present at the interface, distorting but not
vitiating the ordered state. The resulting distortion plays a role in
determining the interfacial energy.  The penetration of force is via a
denatured core of size $\xi$, surrounded by such a distorted region of
size $\lambda$.  These defect blobs could be pinned by certain
sequences, thereby localizing them in specific regions of the DNA. We
speculate that the regions which localize the defect blobs are the
non-coding promoter regions.  This gives a topological interpretation
of the defect blob and it would also be applicable to torque. The
existence of the mixed or Type II phase with pinned defect blobs will
affect the melting profile under a force, and the force-distance
isotherm will show steps originating from the blobs, especially for
finite chains. Our analysis shows that the relation between ordering
and unzipping is needed to get a negative interfacial energy. The
helical ordering is not just base-pairing \textendash \ it involves
stacking and other distant neighbor interactions.  Any microscopic
model for Type II DNA would have to take these into account. On the
experimental front, it is time for a second generation single
molecular experiments that would explore the interfaces on DNA.

To summarize, in this paper we showed that different types of
phenomena happen for two regimes of the ratio
$\kappa=\frac{\lambda}{\xi}$ of the independent length scales $\xi$
and $\lambda$, of DNA order parameter ($\psi$) and internal force
($\gt$) respectively. For $\kappa\ll 1$, the interfacial energy is
positive, and the unzipping or melting under a force is first order.
The external force has no effect inside the ordered, or, zipped phase,
i.e., there is no internal force ($\gt$) inside as $\lambda$ is small.
This is named Type I. On the contrary, for $\kappa\gg~1$, the
interfacial energy becomes negative and the force penetrates the
zipped phase in the form of defect blobs.  The creation of interfaces
are energetically favored, so that interfaces are formed
spontaneously. Thus defect blobs are formed inside the ordered phase.
Above a force threshold $g>g_{c1}$, there will be a finite density of
these defect blobs.  The melting under tension of unzipping is second
order.  This case is named Type II.

\end{document}